\documentclass[]{aastex6}

\usepackage{graphicx}
\usepackage{array}
\usepackage{dcolumn}
\usepackage{color}           

\newcommand{\spacing}[1]{\renewcommand{\baselinestretch}{#1}\large\normalsize}

\begin{document}


\title{Critical Height of the Torus Instability in Two-Ribbon Solar Flares}


\author{Dong Wang\altaffilmark{1,2}, Rui Liu\altaffilmark{1}, Yuming Wang\altaffilmark{1}, Kai Liu \altaffilmark{1}, Jun Chen\altaffilmark{1}, \\ 
Jiajia Liu\altaffilmark{1}, Zhenjun Zhou\altaffilmark{1}, \& Min Zhang\altaffilmark{2}}


\altaffiltext{1}{CAS Key Laboratory of Geospace Environment, Department of Geophysics and Planetary Sciences, University of Science and Technology of China, Hefei, Anhui 230026, China; rliu@ustc.edu.cn}
\altaffiltext{2}{Department of Mathematics and Physics, Anhui Jianzhu University, Hefei 230601, China}


\begin{abstract}

We studied the background field for 60 two-ribbon flares of M-and-above classes during 2011--2015. These flares are categorized into two groups, i.e., \emph{eruptive} and \emph{confined} flares, based on whether a flare is associated with a coronal mass ejection or not. The background field of source active regions is approximated by a potential field extrapolated from the $B_z$ component of vector magnetograms provided by the Helioseismic and Magnetic Imager. We calculated the decay index $n$ of the background field above the flaring polarity inversion line, and defined a critical height $h_\mathrm{crit}$ corresponding to the theoretical threshold ($n_\mathrm{crit}=1.5$) of the torus instability. We found that $h_\mathrm{crit}$ is approximately half of the distance between the centroids of opposite polarities in active regions, and that the distribution of $h_\mathrm{crit}$ is bimodal: it is significantly higher for confined flares than for eruptive ones. The decay index increases monotonously with increasing height for 86\% (84\%) of the eruptive (confined) flares but displays a saddle-like profile for the rest 14\% (16\%), which are found exclusively in active regions of multipolar field configuration. Moreover, $n$ at the saddle bottom is significantly smaller in confined flares than that in eruptive ones. These results highlight the critical role of background field in regulating the eruptive behavior of two-ribbon flares.

\end{abstract}

\keywords {Sun: flares---Sun: coronal mass ejections (CMEs)---Sun: magnetic fields---instabilities}



\section{Introduction} \label{sec:intro}
  
Solar flares and coronal mass ejections (CMEs) are among the most energetic phenomena in the solar system. They are often associated with each other and hence believed to be governed by the same physical process \citep{priest2002magnetic,harrison2003soho,zhang2001temporal,zhang2004study}. In the ``standard'' picture \citep{shibata1998evidence}, a positive feedback is established between the slow rising of a magnetic flux rope and magnetic reconnection underneath; as a result, the flux rope erupts into interplanetary space as a CME, and the reconnection is mapped to the solar surface as two flare ribbons. However, some flares may exhibit circular-shaped \citep[e.g.,][]{Liu2015} or X-shaped ribbons \citep[e.g.,][]{Liu2016SR}, and not all flares are associated with CMEs \citep{Yashiro2005}. Conventionally, flares are categorized as eruptive flares (with CME association) and confined flares (without CME association). \citet{wang2007comparative} suggested that eruptive flares differ from confined ones in both the energy release location and the ratio between magnetic flux in the low ($<$1.1 $R_\sun$) and high ($>$1.1 $R_\sun$) corona. Relevant to the ratio is the torus instability, which has been recognized as a pertinent MHD instability underlying solar eruptions from both theoretical \citep{vanTend&Kuperus1978,kliem2006torus,Aulanier2010} and observational perspectives \citep[e.g.,][]{torok&kliem2005,liu2008magnetic,cheng2011comparative,xu2012relationship,zuccarello2014observational,sun2015great,Liu2016}. The torus instability occurs when the external field above the flux rope decreases too rapidly with increasing height, which is quantified by the decay index $n=-d\ln B/d\ln h$. The threshold value of the instability $n_\mathrm{crit}$ is derived to be 1.5 for a toroidal current channel \citep{kliem2006torus}, while for a very flat, nearly two-dimensional current channel, $n_\mathrm{crit}\gtrsim1$ \citep{Demoulin&Aulanier2010}. On the other hand, some numerical studies \citep[e.g.,][]{Fan&Gibson2007,Kliem2013,zuccarello2016apparent} and laboratory experiments \citep{Myers2015,Myers2016,Myers2017} found that $n_\mathrm{crit}$ is in the range [1.4--2.0]. 

Before the above discrepancy is resolved, we simply take $n_\mathrm{crit}=1.5$ as a yardstick number and define the height corresponding to $n_\mathrm{crit}$ as \emph{critical height} $h_\mathrm{crit}$ to quantify the onset point of the torus instability. We carried out a comprehensive investigation to evaluate in what extent the decay index affects solar eruptions, which has significant implications for space weather forecasting. We selected events from two-ribbon flares occurring during 2011--2015. The working assumption is that a magnetic flux rope is present in a classical two-ribbon flare, no matter the rope is preexistent \citep[e.g.,][]{Liu2010} or newly formed \citep[e.g.,][]{Wang2017}. In the sections that follow we elaborate on the procedure of calculation in \S\ref{sec:calculate} and give the statistical results and concluding remarks in \S\ref{sec:result}. 

\section{Observation \& Analysis}\label{sec:calculate}

\subsection{Instruments}  
This study mainly used data from the Helioseismic and Magnetic Imager \citep[HMI;][]{scherrer2012helioseismic} and the Atmospheric Imaging Assembly \citep[AIA;][]{lemen2011atmospheric} onboard the Solar Dynamics Observatory  \citep[SDO;][]{Pesnell2012}, which was launched on 2010 February 11. HMI's \texttt{hmi.sharp\_cea} data series provide disambiguated vector magnetograms that are deprojected to the heliographic coordinates with a Lambert (cylindrical equal area; CEA) projection method, at a cadence of 720 s and a pixel scale of 0.03$^\circ$ \citep[or 0.36 Mm;][]{Bobra2014}. Flares are monitored by the Geostationary Operational Environmental Satellite (GOES) in soft X-ray (SXR) irradiance and by AIA's seven EUV imaging passbands (94, 131, 171, 193, 211, 304, and 335~{\AA}) and two UV imaging passbands (1600 and 1700~{\AA}) with a spatial resolution of 1.5$''$ and a  temporal cadence of 12~s (24~s) for EUV (UV) passbands \citep {lemen2011atmospheric}. To obtain the context on CMEs, we examined coronagraph images obtained by Solar and Heliospheric Observatory (SOHO) and Solar Terrestrial Relations Observatory (STEREO). 
                 
\subsection{Selection and Category of Events} \label{sec:selection}
60 two-ribbon flares of M- and X-class are selected in this study (Table \ref{tab:mytable}) according to observations of UV flare ribbons in the chromosphere and of EUV post-flare arcades in the corona. The selection criterion is that the center of the source active region is located within $\sim\,$45 degree from the solar disk center, so that the measurements of photospheric magnetic field are relatively reliable. Flares are categorized as either `E' (eruptive) or `C' (confined) in Table \ref{tab:mytable}. To determine whether a flare is associated with a CME, we collated coronagraph images obtained by SOHO and STEREO, and EUV images obtained by AIA. The SOHO LASCO CME catalog\footnote{\url{http://cdaw.gsfc.nasa.gov/CME_list/index.html}} provides a benchmark reference for this purpose. Taking into account the timing and location of flares relative to CMEs as well as the CME speed and direction, we identified 35 eruptive and 25 confined flares (Table \ref{tab:mytable}) .

   \begin{deluxetable}{cccccccccl}
   \spacing{0.5}
   \tablenum{1}
   \tablecaption{Flare list} \label{tab:mytable}
   \tablewidth{0pt}
   \tablehead{
   \colhead{Number} & \colhead{Date} & \colhead{Time\tablenotemark{a}}&
   \multicolumn{2}{c}{Location} & \colhead{Class} & \colhead{Category\tablenotemark{c}} &\colhead{Profile\tablenotemark{d}} &\colhead{Configuration\tablenotemark{e}} & \colhead{$h_\mathrm{crit}$} \\
   \cline{4-5}
   \colhead{} & \colhead{YYYYMMDD} & \colhead{hhmm} & \colhead{AR} & \colhead{Position\tablenotemark{b}}  & \colhead{} & \colhead{} & \colhead{} & \colhead{} & \colhead{(Mm)} 
   }
   \decimals
   \startdata
   1 & 20110307 & 1430 & 11166 & N11E21 & M1.7 & E & I & D &  $46.8 ^{+3.2}_{-3.2}$ \\
   2 & 20110802 & 0619 & 11261 & N17W22 & M1.4 & E & I & M &$22.2 ^{+5.1}_{-6.0}$ \\
   3 & 20111001 & 0959 & 11305 & N10W06 & M1.2 & E & I & M & $24.1 ^{+2.5}_{-1.9}$ \\
   4 & 20111002 & 0050 & 11305 & N09W12 & M3.9 & E & I & M&  $19.4 ^{+2.8}_{+2.3}$ \\
   5 & 20111226 & 0227 & 11387 & S21W33 & M1.5 & E & S & M &$11.6 ^{+3.9}_{+1.6}$  \\
   6 & 20111226 & 2030 & 11387 & S21W44 & M2.3 & E & S &M & $9.3 ^{+3.6}_{-1.9}$  \\
   7 & 20120119 & 1605 & 11402 & N32E27 & M3.2 & E &I & M* & $46.9 ^{+6.0}_{-5.0}$  \\
   8 & 20120123 & 0359 & 11402 & N28W21 & M8.7 & E & I &M* & $46.0 ^{+5.9}_{-6.1}$ \\
   9 & 20120307 & 0024 & 11429 & N18E31 & X5.4 & E & I &D &  $38.6 ^{+2.9}_{-3.4}$ \\
   10& 20120307 & 0114 & 11429  & N15E26  & X1.3 & E & I & D &   $39.1 ^{+7.5}_{-8.1}$  \\
   11& 20120310  & 1744  & 11429  & N17W24  & M8.4 & E & I &D &  $62.4 ^{+10.4}_{-19.9}$  \\
   12& 20120314 &1521  & 11432  &N13E05  & M2.8 & E & I & M &$31.1 ^{+7.8}_{-9.6}$  \\
   13& 20120315 &0752  & 11432 & N14W03 & M1.8 & E & I & M & $40.7 ^{+2.4}_{-3.0}$  \\
   14& 20120606 &2006  & 11494  &S19W05  & M2.1 & E  & I & M* &$20.0 ^{+1.4}_{-1.4}$ \\
   15& 20120614 &1435  & 11504  &S19E06  & M1.9 &  E  & I &D &  $45.8 ^{+4.4}_{-4.7}$  \\
   16& 20120705 &1318  & 11515  &S16W43  & M1.2 & E  & I & M* &$68.9 ^{+5.6}_{-5.2}$  \\
   17& 20120712 & 1649 & 11520  & S13W03 & X1.4 & E  & I & M* &$36.0 ^{+5.0}_{-5.3}$  \\
   18& 20130516 & 2153 &11748  &N11E40  & M1.3 & E  &I & M* &$21.9 ^{+2.4}_{-2.4}$  \\
   19&20130812  &1041  &11817  &S21E18  & M1.5 &E  & I &M* & $22.9 ^{+3.4}_{-2.7}$  \\
   20&20131013  &0043  &11865  &S22E17  & M1.7 &E  & S &M &  $15.3 ^{+7.5}_{-4.3}$  \\
   21& 20131028 & 1153 &11877  &S16W44  & M1.4 &E  & I & D &$69.7 ^{+3.4}_{-3.0}$  \\
   22& 20140212 & 0425 &11974  &S12W02  & M3.7 &E  & S & M &$64.8 ^{+10.6}_{-14.0}$  \\
   23& 20141217 &0110  & 12242 &S20E08  & M1.5 &E  & I & M* &$26.6 ^{+4.3}_{-3.9}$  \\
   24&20141217  &0150  &12241  &S11E33  & M1.1 &E  & I & M &  $15.9 ^{+11.4}_{-3.6}$ \\
   25& 20141217 &0451  &12242  &S18E08  & M8.7 &E  & I & M* & $27.7 ^{+5.7}_{-5.1}$  \\
   26& 20141220 &0028  &12242  &S19W29  & X1.8 &E  & I & M & $40.5 ^{+4.3}_{-4.5}$  \\ 
   27&20141221  &1217  &12241  &S13W25  & M1.0 &E  & I & M & $51.7 ^{+8.2}_{-21.5}$  \\
   28& 20150309 &2353  &12297  &S18E45  & M5.8 &E  & I &M & $31.9 ^{+5.8}_{-6.2}$ \\
   29&  20150315&2322  &12297  &S19W32  & M1.2 &E  & I &M & $ 20.1^{+5.6}_{-5.3}$  \\
   30& 20150316 & 1058 &12297  &S17W38  & M1.6 &E  & S & M & $ 16.8^{+1.9}_{-2.5}$  \\
   31&20150621 & 0142  &12371  &N12E13  & M2.0 &E  & I & D & $46.3 ^{+10.3}_{-12.9}$  \\
   32& 20150622 & 1823 & 12371 & N13W06 & M6.5 & E & I & D & $31.6 ^{+12.3}_{-8.5}$  \\
   33&20150625  &0816 &12371  & N12W40 & M7.9 & E & I & D & $56.6 ^{+5.7}_{-7.4}$  \\
   34& 20151104 &1352 &12443  &N08W02  & M3.7 & E & I & D & $68.7 ^{+3.9}_{-4.4}$  \\
   35& 20151109 &1312 &12449  &S13E39  & M3.9 & E & I & M* &$35.3 ^{+3.3}_{-4.5}$  \\
   36&20110309  &1107 &11166  &N09W06  & M1.7 & C & I & D & $46.9 ^{+7.1}_{-8.5}$  \\
   37&20110803  &0432 & 11261 &N17E12  & M1.7 &C & S &M &  $16.9 ^{+2.0}_{-1.6}$  \\
   38&20111105  &0335 &11339  &N20E45  & M3.7 & C & I & M & $74.5 ^{+8.2}_{-7.6}$  \\
   39&20111105 &1121 & 11339 &N19E41  & M1.1 & C & I & M & $74.9 ^{+8.9}_{-8.3}$  \\
   40& 20111106 & 0103& 11339 &N21E33  & M1.2 & C & I &M &  $82.2 ^{+9.7}_{-9.3}$  \\
   41& 20111231 &1315 &11389  &S25E46  & M2.4 & C & I &M* & $61.4 ^{+1.9}_{-2.3}$  \\
   42& 20111231 &1626 &11389  &S26E42  & M1.5 & C & I &M*& $62.5 ^{+2.5}_{-3.4}$  \\
   43&20120306  &1241 &11429  &N18E36  & M2.1 & C & I & D & $37.9 ^{+5.1}_{-7.6}$  \\
   44& 20120427 &0824 &11466  &N11W30  & M1.0 & C & I &M* & $27.2 ^{+2.1}_{-2.0}$  \\
   45&20120509  &1408 &11476  &N06E22  & M1.8 & C & I & M &$33.6 ^{+4.1}_{-3.8}$  \\
   46& 20120710 &0514 &11520  & S16E35 & M1.7 & C & I &M* & $38.4 ^{+3.6}_{-4.2}$  \\
   47& 20131101 &1953 &11884  &S12E01  & M6.3 & C & S & M* & $71.2 ^{+6.7}_{-7.2}$  \\
   48&  20140204&0400 &11967  &S14W07  & M5.2& C & S & M &$28.4 ^{+7.5}_{-6.6}$ \\
   49&20140206  &2305 &11967  &S15W48  & M1.5 & C & S & M &$19.4 ^{+5.4}_{-3.3}$  \\
   50&20141020  &0911 &12192  & S16E42 & M3.9 & C & I & D & $78.7 ^{+11.8}_{-14.1}$  \\
   51&20141020  &1637 &12192  &S14E39  & M4.5 & C & I & D & $82.1 ^{+12.6}_{-15.2}$  \\
   52& 20141022 &1428 &12192  &S14E13  & X1.6 & C & I & D & $70.6^{+8.4}_{-7.9}$  \\
   53& 20141024 & 2141&12192  &S22W21  & X3.1 & C & I & D & $84.0 ^{+10.5}_{-9.9}$ \\
   54&20141201  &0641 &12222  & S22E17 & M1.8 & C & I & M* &$ 57.1^{+1.6}_{-1.4}$  \\
   55&20141217  &1901 &12241  &S10E23  & M1.4 & C & I & M &  $48.6 ^{+6.6}_{-6.1}$  \\
   56&20141218  & 2158& 12241 &S11E10  & M6.9 & C & I & M & $ 54.7^{+5.6}_{-4.9}$  \\
   57&20141219  &0944 & 12242 &S19W27  & M1.3 & C & I & M* &$48.7 ^{+17.1}_{-9.5}$  \\
   58& 20150103 &0947 &12253  &S05E16  & M1.1 & C & I & D &  $58.9 ^{+1.9}_{-2.5}$  \\
   59& 20150104 & 1536&12253  &S05E01  & M1.3 & C & I & D & $62.5 ^{+0.8}_{-0.8}$  \\
   60& 20150311 &1851 &12297  &S15E18  & M1.0 & C & I & M & $19.6 ^{+5.2}_{-6.4}$  \\
   \enddata
   \tablenotetext{a}{GOES 1--8 {\AA} peak time.}
   \tablenotetext{b}{Flare positiion porvided by GOES.}
   \tablenotetext{c}{`E' for eruptive flares, `C' for confined flares.}
   \tablenotetext{d}{`I' for monotonous increasing of $n$ as a function of $h$, `S' for a saddle-like $n(h)$ profile.}
   \tablenotetext{e}{`D' for a  dipolar magnetic field, `M' for a multipolar field and `M*' indicates that \\
   		the active region of interest is too close to be separated from a neighboring active region. }
   \end{deluxetable}

\subsection{Decay Index \& Critical Height} 
According to an analytical model of torus instability, a toroidal flux ring is unstable to lateral expansion if the external poloidal field $B_{\mathrm{ex}}$ decreases rapidly with increasing height such that the decay index $n=-d\ln B_{\mathrm{ex}}/d\ln h$ exceeds 3/2 \citep{kliem2006torus}. Due to the difficulty in decoupling $B_{\mathrm{ex}}$ from the flux-rope field in either simulation or observation, a conventional practice is to approximate $B_{\mathrm{ex}}$ with a current-free potential field \citep[e.g.,][]{torok2007numerical,Fan&Gibson2007,Demoulin&Aulanier2010,liu2008magnetic}. In our study, the coronal potential field is extrapolated from the $B_z$ component of the vector magnetograms for active regions, using a Fourier transformation method \citep{Alissandrakis1981}. 

\begin{figure}[ht]
\centering
\includegraphics[width=19cm]{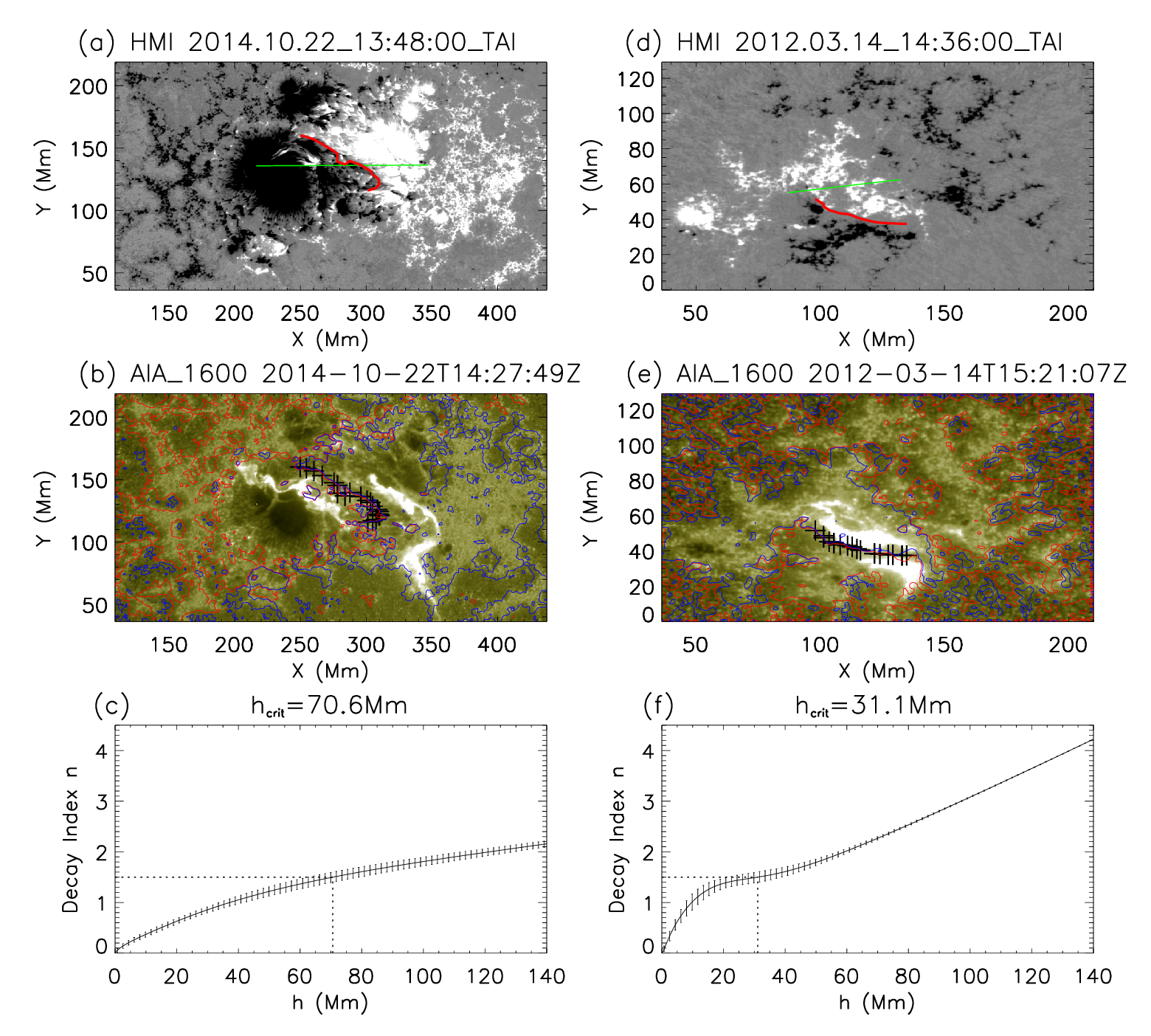}  
\caption{Derivation of the decay index profile for two exemplary events, a confined flare (No.~52) on the left and an eruptive flare (No.~12) on the right. (a) and (d) HMI $ B_z$ map. The red line denotes the flaring PIL and the green line connects the centroids of opposite polarities. (b) and (e) The AIA 1600~{\AA} image overlaid by $ B_z$ contour (50 G and 10 G), with red (blue) colors indicating negative (positive) polarity. The sign `+' denotes the points selected along the flaring PIL. (c) and (f) The decay index $n$ as a function of the height $h$ above the surface in units of Mm. Dotted lines indicate where $n_\mathrm{crit}$ and $h_\mathrm{crit}$ are taken. \label{fig:fig1}}
\end{figure}

Hence in our calculation $n=-d\ln B_t/d\ln h$, where $B_t$ denotes the transverse component of the extrapolated potential field, i.e., $B_t=\sqrt{B_x^2+B_y^2}$. Precisely speaking, it is the external field component orthogonal to the axial current of the flux rope that creates the downward $\mathbf{J}\times\mathbf{B}$ force. $B_t$ often serves as a good approximation since potential field is almost orthogonal to PIL, along which a flux rope in equilibrium typically resides. One needs keep in mind that this approximation works better with less curved PILs. Here we take as an example the confined flare on 2014 October 22 in NOAA AR 12192 \citep[No.\,52 in Table~\ref{tab:mytable}; see also][]{Sun2015,LiuL2016} to demonstrate how the critical height $h_\mathrm{crit}$ is calculated. Figure \ref{fig:fig1}(a) shows a pre-flare photospheric $B_z$ map of AR 12192 at 13:48 UT prior to the onset of the flare and Figure \ref{fig:fig1}(b) the flare ribbons observed near the SXR peak at 14:28 UT in AIA 1600~{\AA}. We sampled the segment of polarity inversion line (PIL) that is located in between the two flare ribbons (referred to as `flaring PIL' hereafter) by clicking on it as uniformly as possible to get sufficient representative points (marked by crosses), and then calculate decay index $n$ at different heights at these selected points. In Figure~\ref{fig:fig1} (c) we plot $n$ as a function of $h$, which is averaged over the selected points, with the error bar indicating the standard deviation. We located the critical height corresponding to $n=1.5$ by linear interpolation between the discrete $n(h)$ points, which have a step of 0.36 Mm, and similarly we located the height at $n=1.5$ on the $n+\delta n$ and $n-\delta n$ profile, where $\delta n$ is the standard deviation at each $n(h)$ point, to get an uncertainty estimation of critical height. For this case, we obtained that $h_\mathrm{crit}=70.6^{+8.4}_{-7.9}$ Mm. As a comparison, Figure~\ref{fig:fig1}(d--f) shows an eruptive flare taking place on 2012 March 14 (No.~12). The corresponding $h_\mathrm{crit}=31.1 ^{+7.8}_{-9.6}$ Mm is much smaller than the confined case.

To evaluate the complexity of magnetic field in active regions and its impact on $h_\mathrm{crit}$, we calculated the centroids of positive and negative magnetic fluxes for each active region and their distance $d$. We propose that the magnetic field relevant to a flare of interest can be deemed as dipolar field (labeled `D' in Table~\ref{tab:mytable}) if the centroids of opposite polarities are located at two sides of, and their connection passes through, the flaring PIL (e.g., Figure~\ref{fig:fig1}a). In contrast, the magnetic field is deemed as multipolar field (labeled `M' in Table~\ref{tab:mytable}) if the connection of centroids fails to pass through (e.g., Figure~\ref{fig:fig1}d), or, is almost parallel to, the flaring PIL. The latter category includes some cases in which the active region of interest cannot be clearly separated from a neighboring active region (labeled `M*' in Table~\ref{tab:mytable}). By visual inspection, we confirmed that this categorization gives a result consistent with the conventional view of dipolar and multipolor field.

\section{Results} \label{sec:result}
\begin{figure}[ht!]
\centering
\includegraphics[height=0.8\textheight]{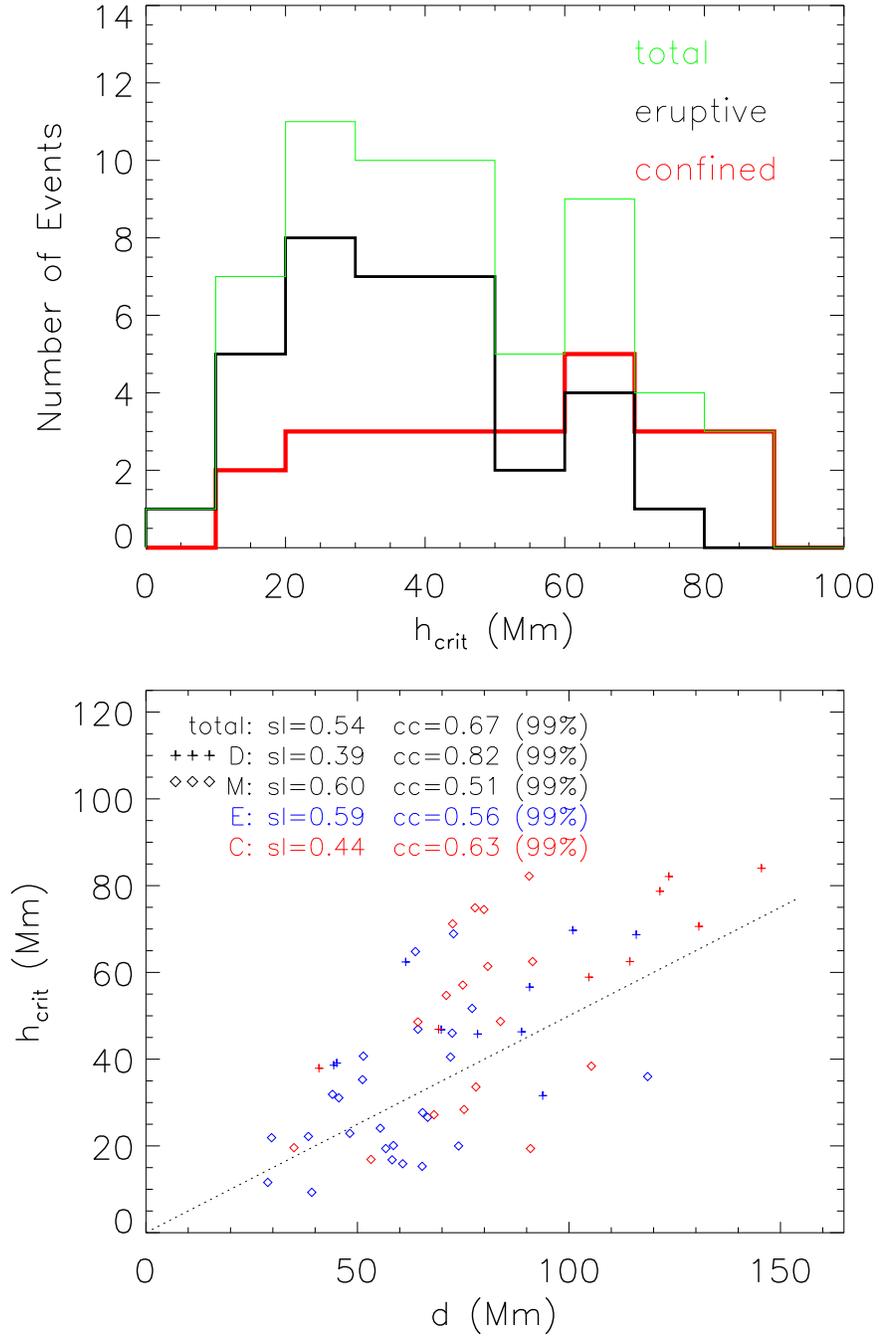}
\caption{Distribution of $h_\mathrm{crit}$ (top) and its relation to the centroid distance $d$ of active regions (bottom). In the bottom panel, plus and diamond symbols denote dipolar (D) and multipolar (M) magnetic field, respectively. Eruptive (`E') and confined (`C') events are shown in blue and red, respectively. `sl' indicates the slope given by linear fitting and `cc' the correlation coefficient with the confidence interval denoted in the brackets. $h_\mathrm{crit}=\frac{1}{2}d$ is marked by the dotted line.   \label{fig:fig2}}
\end{figure}

The distribution of $h_\mathrm{crit}$ for the sample of 60 two-ribbon flares is shown in the top panel of Figure~\ref{fig:fig2}. The total distribution of $h_\mathrm{crit}$ peaks at the heights of 20-30 Mm, but for confined flares $h_\mathrm{crit}$ significantly spreads to higher heights than eruptive flares. The average $h_\mathrm{crit}$ is $36.3\pm17.4$ Mm for the 35 eruptive flares, and $53.6\pm21.3$ Mm for the 25 confined flares. $h_\mathrm{crit}$ is highly correlated with the centroid distance $d$ of active regions (bottom panel of Figure~\ref{fig:fig2}). From the linear fittings using a least absolute deviation method (\texttt{LADFIT} in IDL), we obtained an empirical formula
\begin{equation}
h_\mathrm{crit} \simeq \frac{1}{2}d,  \label{eq:h&d}
\end{equation} 
which may serve as a rule of thumb for the scale of $h_\mathrm{crit}$. In comparison to numerical models, \citet[][Eq.\,15]{Kliem2014} found that within the framework of the active-region model developed by \citet{titov1999basic}, $h_\mathrm{crit}/L$ is slightly below unity, where $L$ is the half distance between two monopoles. This is derived for a freely expanding 
torus without being line-tied. In the numerical experiments with a line-tying surface  \citep[][their Figures 2 and 3]{torok2007numerical}, one can also see that for bipolar configurations $h_\mathrm{crit}$ increases when the distance between external sources increases and that Eq.~\ref{eq:h&d} approximately holds for each case (T.~T\"{o}r\"{o}k, private communication). On the other hand, $h_\mathrm{crit}$ is found to be comparable to the horizontal distance between two sub-photospheric monopoles in a series of numerical simulations imposing different photospheric flows and diffusive coefficients \citep{Aulanier2010,Zuccarello2015,zuccarello2016apparent}. Generally speaking, $h_\mathrm{crit}$ may be affected by various factors including, but not limited to, 1) functional form of the external field; 2) other external sources besides the dipole confining the flux rope; 3) depths of the external sources below the surface. For example, in \citet{torok2007numerical}, the monopoles are very close to the surface, as compared to the significant depths set in \citet{Aulanier2010}. 

\begin{figure}[ht!]
\centering
\includegraphics[height=0.8\textheight]{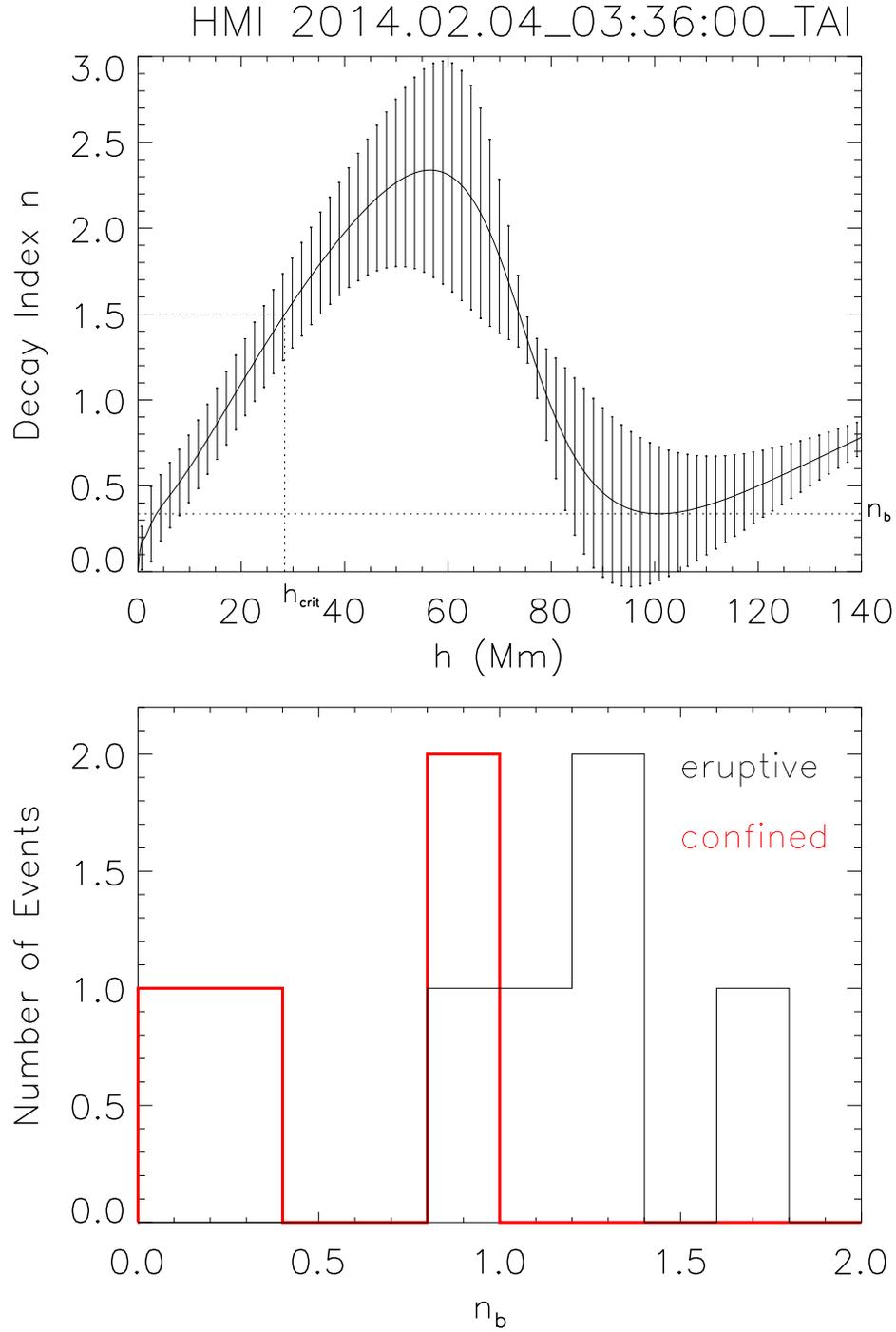}
 \caption{Saddle-like $n(h)$ profile. Top panel shows an exemplary $n(h)$ profile from the confined flare on 2014 February 4 (No.~48 in Table~\ref{tab:mytable}). $n_b$ and $h_\mathrm{crit}$ are marked. Bottom panel shows the distribution of $n_b$ for 5 eruptive (black) and 4 confined (red) flares.  \label{fig:fig3}}
\end{figure}

Two distinct types of $n(h)$ profiles emerge in this investigation, similar to a much smaller sample of 9 flares studied by \citet{cheng2011comparative}: 1) $n$ increases monotonically as the height increases in 30 of 35 (86\%) eruptive flares and in 21 of 25 (84\%) confined flares; and 2) the rest $n(h)$ profiles are saddle-like, exhibiting a local minimum at a height higher than $h_\mathrm{crit}$ (e.g., top panel of Figure~\ref{fig:fig3}). The saddle-like profile provides a potential to confine an eruptive structure if the local minimum $n_b$ at the bottom of the saddle is significantly below $n_\mathrm{crit}$ and the eruption has not developed a large enough disturbance when the eruptive structure reaches the height of $n_b$. For example, the deep saddle bottom at higher altitudes than $h_\mathrm{crit}$ may help confine the eruption on 2014 Feb 4 (top panel of Figure \ref{fig:fig3}). For the 9 flares exhibiting a saddle-like $n(h)$ profile, including 5 eruptive and 4 confined flares, the distribution of $n_b$ is given in the bottom panel Figure \ref{fig:fig3}. One can see that $n_b$ of the 5 eruptive flares (black) is generally larger than that of the 4 confined flares (red). In relation to the field configuration, an outstanding characteristic for saddle-like profiles is that all 9 events originate from multipolar magnetic field (Table~\ref{tbl:stat}).  However, it is not clear exactly what a photospheric flux distribution would yield the saddle shape because, on the one hand, the relevant magnetic field is highly complex; on the other hand, the majority cases of monotonously growing $n(h)$ also originate from multipolar field (Table~\ref{tbl:stat}). This will be considered in a future investigation. 

\setcounter{table}{1}
\begin{table}[ht!]
	\caption{The number of different type of flares and $n(h)$ profiles in relation to magnetic field configuration of active regions. The same notations are adopted here as in Table~\ref{tab:mytable}. \label{tbl:stat}} 
	\[
	\begin{array}{c|cccc}
	& I & S & E & C\\ \hline
	D & 18 & 0 & 10 & 8 \\
	M (M^*) & 33\,(16) & 9\,(1) & 25\,(10) & 17\,(7)
	\end{array} \]
	
\end{table}

To conclude, this investigation confirms that the decay index profile of the background field plays an important role in deciding whether a two-ribbon flare would lead up to a CME. Moreover, the saddle-like profile present in some active regions may provide an additional confinement effect on eruptions. These results indicate the possibility that some two-ribbon flares might be innately incapable of producing CMEs.  

\acknowledgments The authors thank the anonymous reviewer for constructive comments that help greatly improve the paper. D.W. acknowledges the support by Natural Science Foundation of Anhui province education department (KJ2016JD18, KJ2017A493). R.L. acknowledges the support by NSFC 41474151 and the Thousand Young Talents Program of China, and thanks T. T\"{o}r\"{o}k for helpful comments. Y.W. acknowledges the support from NSFC 41131065 and 41574165. This work was also supported by NSFC 41421063, CAS Key Research Program KZZD-EW-01-4, and the fundamental research funds for the central universities.


\end{document}